# Cooperative Couplings between Octahedral Rotations and Ferroelectricity in Perovskites

Teng Gu[1,2], Timothy Scarbrough[3], Yurong Yang[3], Jorge Íñiguez[4], L. Bellaiche[3*], and H. J. Xiang[1,2,3*]

[1]*Key Laboratory of Computational Physical Sciences (Ministry of Education), State Key Laboratory of Surface Physics, and Department of Physics, Fudan University, Shanghai 200433, P. R. China*

[2]*Collaborative Innovation Center of Advanced Microstructures, Nanjing 210093, P. R. China*

[3]*Physics Department and Institute for Nanoscience and Engineering, University of Arkansas, Fayetteville, Arkansas 72701, USA*

[4]*Materials Research and Technology Department, Luxembourg Institute of Science and Technology (LIST), 41 Rue du Brill, L-4422 Belvaux, Luxembourg*

*email: laurent@uark.edu, hxiang@fudan.edu.cn

**Abstract**

The structure of $ABO_3$ perovskites is dominated by two types of unstable modes, namely, the oxygen octahedral rotation (AFD) and ferroelectric (FE) mode. It is generally believed that such AFD and FE modes tend to compete and suppress each other. Here we use first-principles methods to show that a dual nature of the AFD—FE coupling, which turns from competitive to cooperative as the AFD mode strengthens, occurs in numerous perovskite oxides. We provide a unified model of such a dual interaction by introducing novel high-order coupling terms, and explain the atomistic origin of the resulting new form of ferroelectricity in terms of universal steric mechanisms. We also predict that such a novel form of ferroelectricity leads to atypical behaviors, such as an enhancement of the electric polarization under hydrostatic pressure.

Ferroelectricity is a material property to have a spontaneous electric polarization that can be reversed by an external electric field. Perovskite-based ferroelectrics (e.g., $BaTiO_3$, $PbTiO_3$, $BiFeO_3$) are the best known and studied ferroelectrics, not only because they are important for applications, but also because their simple structure makes them ideal for understanding the origin of this effect. In fact, it was demonstrated[1] that the hybridization between the low-lying empty d states of the B-ion (e.g., $Ti^{4+}$) and the O 2p states (the so-called second-order Jahn–Teller effect[2]) and/or the interaction between lone-pairs of A-ion (e.g., $Pb^{4+}$ and $Bi^{3+}$) and O 2p states are key factors for the occurrence of ferroelectricity in proper $ABO_3$ perovskite oxides. It is now therefore widely believed that the presence of one of these two features, or both of them, is essential for proper ferroelectricity to happen in $ABO_3$ compounds.

This common belief is, in fact, challenged by the recent experimental discoveries that many $ABO_3$ with unusually small A-ions are polar and/or ferroelectric (FE). For instance, Inaguma *et al.* reported[3] the synthesis of $ZnSnO_3$ with apolar R3c $LiNbO_3$-type (LN-type) structure, while Sn has no low-lying empty d states and Zn does not possess any lone pair. Subsequently, $ZnSnO_3$ thin films[4] were further shown to exhibit a rather high and switchable ferroelectric polarization of ~47 $\mu C/cm^2$. Moreover, the calculated Born effective charges[5] for $ZnSnO_3$ are close to the corresponding nominal charges, confirming that the conventional covalent interaction mechanism is not responsible for ferroelectricity in $ZnSnO_3$. Strikingly, other LN-type materials with small A ions, such as $ZnPbO_3$[6,7], $ScFeO_3$[8,9], $InFeO_3$[10], $PbNiO_3$[11] were synthesized as well. Some of them (e.g., $Mn_2FeTaO_6$, $Zn_2FeOsO_6$)[12-14] even simultaneously adopt ordered magnetism, which therefore renders them multiferroic. In these materials, A-site driven FE distortions coexist with large antiferrodistortive (AFD) modes (rotations/tilts of the $O_6$ octahedra)[15]. However, those structural distortions are known to typically compete in perovskites[16-19]. Since these ferroelectrics are promising for the generation of Pb-free ferroelectric thin films and the realization of room-temperature multiferroics, it is highly desirable to understand the mechanism of their ferroelectricity, and why/how the strong FE and large AFD modescan coexist.

In this Letter, we perform density functional theory (DFT) simulations [see section 1 of the Supplementary Materials (SM)]to reveal that the FE—AFD couplings follow a universal behavior in many perovskite oxides, changing from competitive to cooperative as the $O_6$ tilts increase. In particular, these couplings are found to be cooperative in the regime that is relevant to LN-type compounds like $ZnSnO_3$, and explain the ferroelectricdistortion in this and related materials(e.g.,the "ferroelectric" metal $LiOsO_3$[20]).We further provide an effective Landau-type theory and discuss the underlying atomistic mechanism to find that the cooperative FE—AFD coupling originates from a general steric effect.

Figure 1csummarizes our most important result, revealing the dual nature of the FE—AFD couplings for diverse $ABO_3$ perovskite compoundssuch as$ZnSnO_3$, $BiFeO_3$, $InFeO_3$, $ScFeO_3$and $LiOsO_3$. The rhombohedral (non-centrosymmetric) R3c ground states of these compounds can be formedby superimposing two different distortions onto the ideal cubic perovskite structure, as shown in Figs. 1a and 1b: (i) the FE mode that consists of the collective motions of A and B ions along the pseudo-cubic [111] direction, in addition to O ions moving overall along the opposite [-1-1-1]direction; and (ii) antiphase oxygen octahedral rotations about the pseudo-cubic [111] direction (the so-called $a^-a^-a^-$ pattern in Glazer'snotation)[21]. To produce Fig. 1c, we computed the phonon spectrum [see Fig. S1 of SMas a typical example]in the R-3c phaseswhich are obtained by superimposing the $a^-a^-a^-$tilt (with different amplitude defined with a reduced dimensionless unit [22]) to the ideal cubic perovskite structure[23]. We always observe the same pattern: For relatively small tilts, the frequency of the FE modealong the [111] pseudo-cubic direction typically grows and approaches zero, implying that the polar instability gets weaker, as consistent with the fact that octahedral tilts have been reported to suppress ferroelectricity in many perovskites (see, e.g., Refs. [15,16,24-26] and references therein). However, the tendency reverts for relatively large tilts: indeed, we find that, as the tilting amplitude grows, we obtain more negative frequencies and, correspondingly, stronger FE instabilities. We also examined the effect of a ferroelectrically active B-ion (e.g, $Ti^{4+}$ ion, for a detailed discussion, see section 3 of SM) to find that the FE—AFD coupling is also cooperative there when the tilt is large. Hence, we find that the FE—AFD competition turns into a strong cooperation for large-enough tilt amplitudes for all investigated compounds (i.e., those examined in Fig. 1c and Fig. S2) with tolerance factors ranging from 0.78 to 1.00(section 4 of SM).

Having established our basic result, we now focus on a more detailed discussion on $ZnSnO_3$, which is a good representative material to investigate LN-type compounds. Before we move ahead, let us note that, while we fully realize that the LN-type and perovskite structures are qualitatively different, here we will adopt the approach that LN-type compounds can be thought of as perovskites with very small tolerance factors and large AFD distortions. This will be useful for a unified discussion of the relevant couplings, and to emphasize their observed generality.

Within a reduced dimensionless unit[22], the amplitudes of the FE mode and oxygen octahedral rotation (R) in the R3c ground state of $ZnSnO_3$ are numerically found to be 0.21 and 0.49, respectively. Correspondingly, we obtain the electric polarization to be 55 $\mu C/cm^2$, in agreement with previous experimental[4] (47 $\mu C/cm^2$) and theoretical[5] (57 $\mu C/cm^2$) results, and the rotational angle θ of the oxygen octahedra is as large as ~19° (Note that $\tan(\theta) = \frac{\sqrt{2}}{2}R$).

To further investigate the dual nature of the FE—AFD coupling, we calculate the FE amplitude resulting in the lowest energy for a given rotation amplitude for R3c $ZnSnO_3$. The dependencies of such FE amplitude and corresponding total energy on the octahedral rotation amplitude are both shown in Fig. 2a (similar results are found for other systems, see section 5 of SM).Three different structural relaxation strategies are adopted, in order to determine the influence of the lattice vectors on the FE amplitude and total energy: they are what we denote here as ``fix cell'', ``fix cell shape'' and ``relax cell''. In each of these three cases, the FE mode amplitude is optimized during the structural relaxation while the octahedral rotation amplitude is kept fixed at a given value.These three relaxation schemes differ in the way the cell is relaxed: in the ``fix cell'' scheme, the lattice vectors are fixed to that of the paraelectric cubic Pm-3mstructure as obtained from a symmetry-constrained relaxation of ZnSnO$_3$; in the ``fix cell shape'' case, the cell is still chosen to be cubic (thus implying that no rhombohedral straincan exist) but its volume can relax; in the ``relax cell'' scenario, the lattice vectors are fully optimized. Several interesting trends can be observed from Fig. 2a: (1) the total energy has a minimum at about R = 0.5 for all these three different structural relaxation schemes, which is close to the equilibrium octahedral rotation amplitude in the fully relaxed R3c ground state of ZnSnO$_3$;(2) the amplitude of the FE mode first decreases and then increases with R, which is inline with the dependence of the phonon frequency on the rotationalamplitudedepicted in Fig. 1c. Such non-monotonic behavioroccurs for allthree considered elastic constraints, therefore indicating that *strain relaxations do not play any qualitatively role* in the observed the dual nature of the coupling between FE and oxygen octahedral rotation modes, in sharp contrast with the interesting finding for tetragonal SrTiO$_3$ by Aschauer and Spaldin[27]; (3) the main difference between the different relaxation schemes is that the minimum of the FE mode locates at different rotation amplitudes. In particular, if more strain degrees of freedom are allowed to relax (e.g., if one goes from ``fix cell'' to ``relax cell''), the rotation amplitude associated with the minimal value of the FE amplitudebecomes smaller.It is also interesting to realize that in the region corresponding to the lowest energies (i.e., for AFD amplitudes around 0.5), the strength of the FE mode increases with R in the ``relax cell'' situation while it adopts an opposite behavior in the ``fix cell'' scenario.

In order to better understand the dual nature of the FE-AFD coupling in R3c ZnSnO$_3$, we nowintroduce a Landau-like potential. More precisely, using group theory allows us to derive the following energy expression (up to sixth and fourth orders in FEand AFD mode amplitudes, respectively) for the distorted R3c perovskite structure with respect to the cubic paraelectric state:

$$E(u,R) = A_2(R)u^2 + A_4(R)u^4 + A_6(R)u^6 + E(R) \qquad \text{Eq. (1)},$$

where $u$ and $R$ represent the amplitudes of the FE and AFD modes, respectively.

$E(R)$ characterizes the pure rotational part of energy. $A_2(R) = A_{u^2} + A_{u^2R^2}R^2 + A_{u^2R^4}R^4$, $A_4(R) = A_{u^4} + A_{u^4R^2}R^2 + A_{u^4R^4}R^4$ and $A_6(R) = A_{u^6} + A_{u^6R^2}R^2$, with $A_{u^2R^2}$, $A_{u^2R^4}$, $A_{u^4R^2}$, $A_{u^4R^4}$ and $A_{u^6R^2}$ being coefficients quantifying different couplings between the FE and AFD modes. Note that the strain degree of freedom is not explicitly included in this Landau potential, for simplicity and because we showed above that strain does not qualitatively change the dual nature of the FE—AFD coupling. We find that this Landau potential can describe well the interactions between FE and AFD modes. To demonstrate that, we first fix the lattice vectors to be those of the fully relaxed cubic paraelectric Pm-3m state and report in Fig. 2b the total energy of the R3c phase as a function of the FE mode amplitude for three different, fixed R values: (1) R = 0.4, which falls into the region for which the rotational mode typically suppresses the FE mode (see Fig. 2a); (2) R = 0.5, near which the FE mode has a minimum for the ``fixed cell'' case of Fig. 2a; and (3) R = 0.6, for which the AFD mode has the tendency to enhance the FE mode (see Fig. 2a). For each of these three considered R values, the fitted curve (dashed line) of Fig. 2b corresponds to the proposed Landau potential and matches very well with the calculated DFT results (shown by means of symbols), therefore supporting the suitability of the model. It can be seen in the inset of Fig. 2b that the resulting $A_2$ parameter (as extracted from the fit of DFT calculations with many different, fixed R amplitudes and frozen lattice vectors corresponding to the relaxed cubic paraelectric state) first increases but then decreases with R, while always remaining negative. This naturally implies that $A_{u^2R^2} > 0$ and $A_{u^2R^4} < 0$. The existence of the two couplings $A_{u^2R^2}$ and $A_{u^2R^4}$ having opposite signs evidences the dual nature of the FE—AFD coupling in ZnSnO$_3$: positive $A_{u^2R^2}$ is responsible for the competition between these two modes for smaller R, while negative $A_{u^2R^4}$ testifies of the collaborative nature of the FE and AFD modes for larger R. Note that, if the $u^2R^4$ term is excluded from the Landau model, the FE mode would be greatly suppressed by the tilt: in fact, we numerically found that there would be no FE instability for R = 0.4, 0.5, and 0.6 in that case (see solid lines of Fig. 2b), which demonstrates the crucial role that this previously overlooked $u^2R^4$ coupling plays on the onset of ferroelectricity in compounds like $ZnSnO_3$.

To shed further light on this dual nature, we compute force-constant matrices of R-3c $ZnSnO_3$ with a given fixed rotation amplitude (with the cell being fixed to be that of relaxed cubic structure of $ZnSnO_3$). In particular, we plot the self-force constants of the Zn ion ($\Phi^{self}_{\|[111],\|[111]}$ and $\Phi^{self}_{\perp[111],\perp[111]}$ [28]) as a function of the octahedral rotation amplitude in Fig. 3a. Note that $\Phi^{self}_{\perp[111],\perp[111]}$ always increases with R, which implies that the $a^-a^-a^-$ rotation always tends to suppress any component of the electrical polarization that is perpendicular to the pseudo-cubic [111] direction. On the other hand, $\Phi^{self}_{\|[111],\|[111]}$ first increases slowly with the rotation amplitude, but then decreases rapidly towards much strongly negative values when R goes above ~0.5. Such results indicate that small $a^-a^-a^-$ rotations tend to suppress electrical polarization along the [111] direction, while large $a^-a^-a^-$ rotations tend to enhance it. Let us recall that a negative value of the Zn self-force constant $\Phi^{self}_{\|[111],\|[111]}$ is the landmark of an unstable phonon band dominated by the Zn off-centering (polar) motions, a feature that can be expected in LN-like compounds.

To find out which precise microscopic atomic interaction is responsible for the dependence of the self-force constant of the Zn ion on oxygen octahedral rotation, we further analyze the local environment of the Zn ion in detail. In the relaxed cubic structure, each Zn ion has twelve nearest-neighboring (NN) oxygen ions (see Fig. 3b), leading to twelve symmetry-equivalent Zn-O bonds of length 2.85Å. When the $a^-a^-a^-$ rotation takes place, and assuming that the FE mode continues to have zero amplitude within the relaxed cubic cell, these twelve NN oxygen ions split into three groups: (1) three oxygens lie on the (111) plane containing the Zn ion and get closer to this Zn ion; (2) three additional oxygens lie on the same (111) plane, but separate from the central Zn ion; and (3) six out-of-plane oxygens that separate only slightly from the central Zn ion. For instance, for the two non-vanishing octahedral rotation amplitudes shown in Fig.3b (i.e., R=0.35 and R=0.69), the shortest Zn-O distances are 2.28Å and 1.71Å, respectively, to be compared with the ``expected'' length of the Zn-O bond of 2.09 Å -- as derived by summing the Shannon's[29] ionic radii of six-fold coordinated $Zn^{2+}$ (which is 0.74 Å) and of two-fold coordinated $O^{2-}$ (1.35Å). Thus, when R is small (e.g., R = 0.35), the shortest Zn-O distance is larger than the expected Zn-O bond length. In contrast, the shortest Zn-O distance is smaller than the expected Zn-O bond length when R is large (e.g., R = 0.69). This clearly suggests that the dual nature of the FE—AFD coupling has a simple steric origin. In fact, the interaction between the $Zn^{2+}$ and $O^{2-}$ ions includes two contributions, namely, an attractive one (electrostatic or covalent bonding) and a repulsive one (Pauli repulsion). When the octahedral rotation amplitude is small (i.e., R smaller than 0.5), the attractive part of potential between Zn and O dominates and the three NN oxygen ions attract the Zn ion and tend to keep it in plane. In contrast, when the octahedral rotation amplitude is large enough (e.g., R=0.69), the shortest NN Zn-O distance is smaller than the optimal Zn-O bond length, which results in a repulsive force. The three NN oxygen ions consequently push the central Zn ion out of plane, thus lowering the repulsive contribution to the energy and inducing an electric polarization along the pseudo-cubic [111] direction. Therefore, a size effect leads to the non-monotonic dependence of the Zn self-interaction force constant $\Phi^{self}_{\|[111],\|[111]}$ on the AFD amplitude. Our above explanation can also be validated with a simple Lennard-Jones (LJ) like model potential (section 6 of SM).

Note that Benedek and Fennie[15] also noticed the local environment of the Zn ion in the R-3c structure of $ZnSnO_3$. They proposed that the driving force for the FE instability in R-3c structure can be understood from bond valence arguments. However, they did not appreciate the collaborative effect between AFD and FE modes to explain ferroelectricity in $ZnSnO_3$. In fact, they used a Landau-like model only up to fourth order, which excludes the presence of the $u^2R^4$ term required to reproduce the dual-coupling behavior that we observe from our simulations. We further show that the FE-AFD dual coupling also exists for tilt patterns other than the $a^-a^-a^-$ tilt (section 7 of

SM).A recent theoretical work[27] on $SrTiO_3$ suggests that FE and AFD modes may actually cooperate because of an effect mediated by strain if one were able to access a suitable regime. However, the *steric*effect (not necessarily involving strain) responsible for the FE-AFD cooperation is more general than and different from the special *strain* mechanism for $SrTiO_3$ (section 8of SM). The novel FE—AFD cooperative coupling revealed in this work also leads to atypical effects. For example, ferroelectricity is enhanced by hydrostatic pressure in the R3c phase of $ZnSnO_3$ (section 9of SM).

In summary, we reveal the dual nature of the FE-AFD coupling in many $ABO_3$ perovskites.The ferroelectricity in perovskites with small tolerance factors can be viewed as driven by a FE—AFD cooperative coupling. We provide a unified model of such a dual interaction by introducing novel high-order coupling terms in a Landau-like theory, and explain the atomistic origin of the resulting new form of ferroelectricity in terms of universal steric or size mechanisms. Further, we dispute the common-knowledge that, in these materials, FE—AFD couplings are always competitivein any regime of practical importance.Indeed, we show that, while the large-AFD range is not relevant for some common perovskites (like e.g. $CaTiO_3$ or $SrTiO_3$), it does apply to all-important recently synthesized$LiNbO_3$-like compounds (like e.g. $ZnSnO_3$ or $LiOsO_3$) that display a non-centrosymmetric ground state combining AFD and FE distortions.

**Acknowledgments**

Work at Fudan is supported by NSFC (11374056), the Special Funds for Major State Basic Research (2015CB921700), Program for Professor of Special Appointment (Eastern Scholar), Qing Nian Ba Jian Program, and Fok Ying Tung Education Foundation. H.X. and T.S. also thank the support of the Department of Energy, Office of Basic Energy Sciences, under contract ER-46612, and Y.Y. and L.B. acknowledge the ONR Grant N00014-17-1-2818and ARO Grant No. W911NF-16-1-0227, respectively. We also acknowledge funding from the Luxembourg National Research Fund through the inter-mobility (Grant 15/9890527 GREENOX, J.I. and L.B.) and

Pearl (Grant P12/4853155 COFERMAT, J.I.) programs.

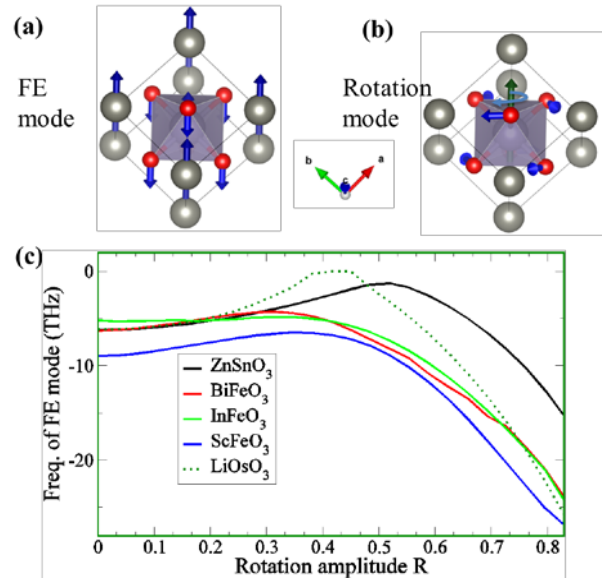

Fig.1 Schematization of the (a) FE and (b) rotation modes, respectively, and (c) dependence of the frequency of the (unstable) FE mode on the rotation mode in the R-3c phase of $ABO_3$.

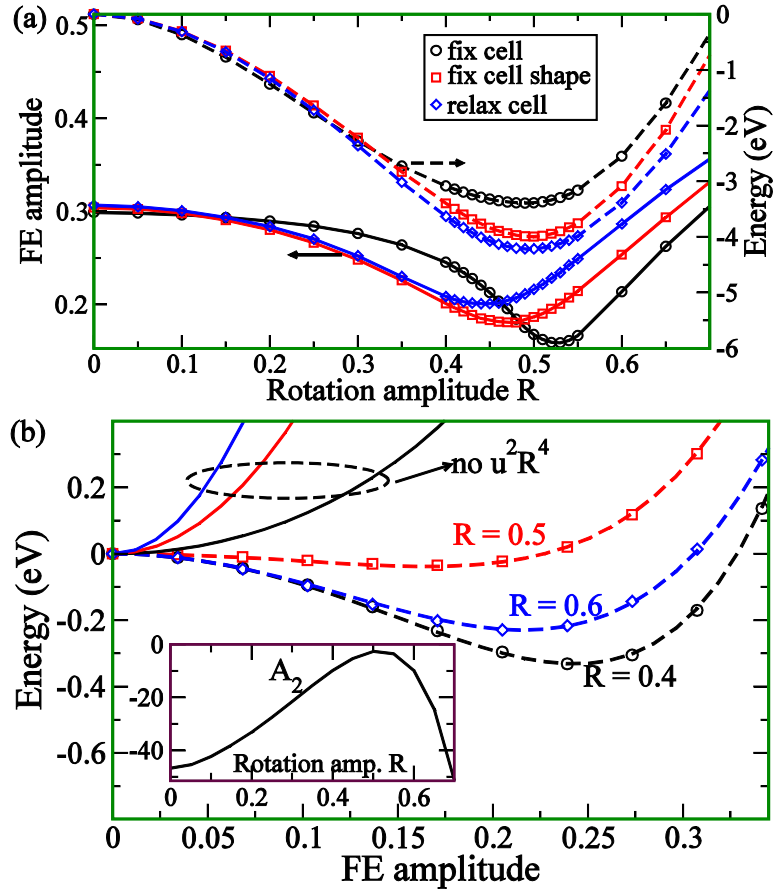

Fig.2 Couplings between the rotation and FE modes in the R3c state of $ZnSnO_3$. (a) FE amplitude and total energy versus the amplitude of the rotation mode. Three different structural relaxation strategies are chosen (i.e., ``fix cell'', ``fix cell shape'' and ``relax cell''), as explained in the text. (b) Total energy (in eV/10-atoms) as a function of the FE amplitude when the rotation mode is fixed and when choosing the lattice vectors of the cubic relaxed Pm-3m state of $ZnSnO_3$. Three different rotation amplitudes (R = 0.4, 0.5, 0.6) are chosen. Symbols and dashed line represent DFT results and their fit by Eq. (1), respectively. The solid lines are the model results excluding the $u^2R^4$ term. The inset in Panel (b) shows the resulting fitted $A_2$ parameter (in eV/10-atoms) as a function of the rotational amplitude.

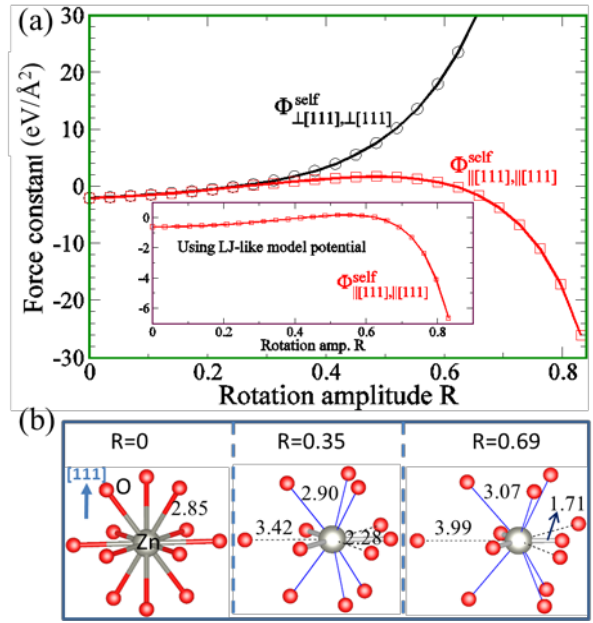

Fig.3 Microscopic origin of the dual nature of the coupling between the rotation and FE modes. (a) The self-force constants of the Zn ion as a function of the octahedral rotation amplitude in R-3c ZnSnO$_3$. Two cases ($\Phi^{self}_{\perp[111],\perp[111]}$ and $\Phi^{self}_{\parallel[111],\parallel[111]}$) are shown, which represent the force constants perpendicular to the [111]-direction or along the [111]-direction, respectively. (b) Local environment around the Zn ion for three different rotation amplitudes (R = 0, 0.35 and 0.69). The Zn-O distances (in Å) are indicated.

**References**

[1]  R. E. Cohen, Nature **358**, 136 (1992).
[2]  J. M. Rondinelli, A. S. Eidelson, and N. A. Spaldin, Physical Review B **79**, 205119 (2009).
[3]  Y. Inaguma, M. Yoshida, and T. Katsumata, Journal of the American Chemical Society **130**, 6704 (2008).
[4]  J. Y. Son, G. Lee, M.-H. Jo, H. Kim, H. M. Jang, and Y.-H. Shin, Journal of the American Chemical Society **131**, 8386 (2009).
[5]  M. Nakayama, M. Nogami, M. Yoshida, T. Katsumata, and Y. Inaguma, Advanced Materials **22**, 2579 (2010).
[6]  R. Yu, H. Hojo, T. Mizoguchi, and M. Azuma, Journal of Applied Physics **118**, 094103 (2015).
[7]  D. Mori, K. Tanaka, H. Saitoh, T. Kikegawa, and Y. Inaguma, Inorganic Chemistry **54**, 11405 (2015).
[8]  M.-R. Li *et al.*, Journal of the American Chemical Society **134**, 3737 (2012).
[9]  T. Kawamoto *et al.*, Journal of the American Chemical Society **136**, 15291 (2014).

[10] K. Fujita *et al.*, Chemistry of Materials **28**, 6644 (2016).

[11] Y. Inaguma, K. Tanaka, T. Tsuchiya, D. Mori, T. Katsumata, T. Ohba, K.-i. Hiraki, T. Takahashi, and H. Saitoh, Journal of the American Chemical Society **133**, 16920 (2011).

[12] M.-R. Li, P. W. Stephens, M. Retuerto, T. Sarkar, C. P. Grams, J. Hemberger, M. C. Croft, D. Walker, and M. Greenblatt, Journal of the American Chemical Society **136**, 8508 (2014).

[13] M.-R. Li *et al.*, Angewandte Chemie International Edition **52**, 8406 (2013).

[14] P. S. Wang, W. Ren, L. Bellaiche, and H. J. Xiang, Physical Review Letters **114**, 147204 (2015).

[15] N. A. Benedek and C. J. Fennie, The Journal of Physical Chemistry C **117**, 13339 (2013).

[16] I. A. Kornev, L. Bellaiche, P. E. Janolin, B. Dkhil, and E. Suard, Physical Review Letters **97**, 157601 (2006).

[17] W. Zhong and D. Vanderbilt, Physical Review Letters **74**, 2587 (1995).

[18] N. Sai and D. Vanderbilt, Physical Review B **62**, 13942 (2000).

[19] J. C. Wojdeł, P. Hermet, M. P. Ljungberg, P. Ghosez, and J. Íñiguez, Journal of Physics: Condensed Matter **25**, 305401 (2013).

[20] Y. Shi *et al.*, Nat Mater **12**, 1024 (2013).

[21] A. M. Glazer, Acta Crystallographica Section B **28**, 3384 (1972).

[22] The magnitude of the mode is defined to be 1 if the sum of the square of the displacements equal to the square of the cubic lattice constant.

[23] We first optimize the lattice constants of the ideal cubic (Pm-3m) perovskite structures for each of the considered materials.

[24] W. Zhong, D. Vanderbilt, and K. M. Rabe, Physical Review B **52**, 6301 (1995).

[25] D. Vanderbilt and W. Zhong, Ferroelectrics **206**, 181 (1998).

[26] S. Amisi, E. Bousquet, K. Katcho, and P. Ghosez, Physical Review B **85**, 064112 (2012).

[27] U. Aschauer and N. A. Spaldin, Journal of Physics: Condensed Matter **26**, 122203 (2014).

[28] Here, we focus on two particular force constants related to the Zn cations, namely $\Phi^{self}_{\|[111],\|[111]}$ and $\Phi^{self}_{\perp[111],\perp[111]}$, where ``self'' indicates that this is the self-interaction of a single Zn ion, while $\|[111]$ and $\perp[111]$ indicate the motion of this Zn ion parallel and perpendicular to the pseudo-cubic [111] direction, respectively.

[29] R. D. Shannon, Acta Crystallographica Section A **32**, 751 (1976).

Supplementary Materials for

# Cooperative Couplings between Octahedral Rotations and Ferroelectricity in Perovskites

Teng Gu[1,2], Timothy Scarbrough[3], Yurong Yang[3], Jorge Íñiguez[4], L. Bellaiche[3*], and H. J. Xiang[1,2,3*]

[1]*Key Laboratory of Computational Physical Sciences (Ministry of Education), State Key Laboratory of Surface Physics, and Department of Physics, Fudan University, Shanghai 200433, P. R. China*

[2]*Collaborative Innovation Center of Advanced Microstructures, Nanjing 210093, P. R. China*

[3]*Physics Department and Institute for Nanoscience and Engineering, University of Arkansas, Fayetteville, Arkansas 72701, USA*

[4]*Materials Research and Technology Department, Luxembourg Institute of Science and Technology (LIST), 41 Rue du Brill, L-4422 Belvaux, Luxembourg*

The goal of this Supplementary material (SM) is toprovide additional information aboutthe method we used as well as our findings.

**1. Computational details**

The density functionaltheory (DFT) method is used for structural relaxationand electronic structure calculations. The ion-electron interactionis treated by the projector augmented wave (PWA) method[1,2]as implemented in the Vienna *ab initio* simulation package (VASP) [3,4]. The exchange-correlation potential is treated by PBE[5]. The plane-wave cutoff energy is set to 500 eV, and all the atoms are allowed to relax untilatomic forces are smaller than 0.01 eV/Å on any ion. The electric polarization is computed with the Berry phase method[6-8].For $ATiO_3$ (A= Ca, Sr, Ba), we adopted the same framework as that used by Aschauer and Spaldin data[9] for comparison. In particular, the PBEsol[10] exchange-correlation functional is used.

For the calculationof the force-constant and phonon spectrum, we use the finitedisplacement method[11] as implemented in the PHONOPY code[12].

Note also that, in order to obtain the FE mode which leads to the lowest energy for a given rotation amplitude, we modified the VASP code to optimize the amplitude of the FE mode. This is achieved by keeping only the atomic forces that are compatible with the FE mode displacements.

## 2. Phonon spectrum of R-3c ZnSnO$_3$

Figure S1 provides the full phonon spectra of the R-3c state of ZnSnO$_3$ with different magnitudes of the rotation, R. We find that (1) the frequency of the FE mode is always negative for any considered R (including R=0), implying that the ferroelectricity in ZnSnO$_3$ is of proper nature; and (2)the frequency of the FE mode first increases with the rotation amplitude, then decreases with the rotation amplitude, revealing a dual nature of the interaction between the AFD and FE modes.

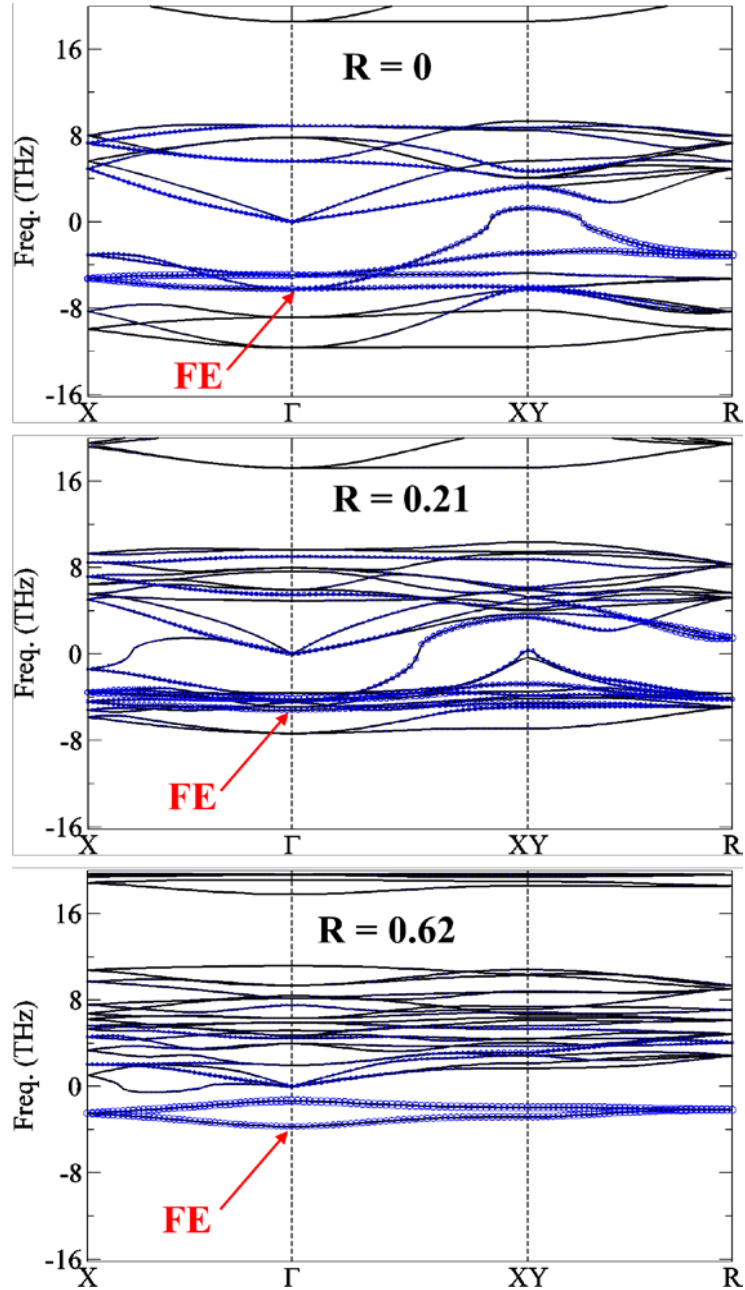

Figure S1. Phonon spectra of the 10-atom R-3c phases of $ZnSnO_3$. The FE mode isindicated via an arrow. The Zn contributions to phonon modes are representedby means of blue circles, indicating that the FE mode is mostly contributed by the displacements of the Zn ions.Three rotational amplitudes (R=0, 0.21 and 0.62, respectively) are considered here.

## 3. Dependence of the frequency of the FE modes on the rotation modes in $ATiO_3$ bulk compounds

In the manuscript, we mainly focus on $ABO_3$ perovskite oxides where the B site does not contain any low-lying empty d orbitals. Here, we examine the coupling between the FE mode and tilts in ATiO$_3$ (A = Ca, Sr, Ba). Since the Ti$^{4+}$ ion is a pseudo-Jahn-Teller active ion, BaTiO$_3$ is ferroelectric while SrTiO$_3$ is known to be an incipient ferroelectric[13]. In contrast, CaTiO$_3$ with a relative small tolerance factor is not ferroelectric because the strong $a^-a^-c^+$ tilt suppresses ferroelectricity (while favoring antipolar motions[14]). For the $a^-a^-a^-$ tilt, the frequency of the FE mode first increases slowly with the rotation amplitude R when R < 0.2, then it decreases with the rotation amplitude R in CaTiO$_3$ (see Fig. S2). Thus, there is also a dual coupling between the FE mode and $a^-a^-a^-$ tilt in R-3c CaTiO$_3$. For SrTiO$_3$ and BaTiO$_3$ (also shown in Fig. S2), the frequency of the FE mode always decreases with the rotation amplitude of the $a^-a^-a^-$ tilt. This suggests that there is a surprising cooperative coupling between the FE mode and $a^-a^-a^-$ tilt in R-3c SrTiO$_3$ and BaTiO$_3$, even for small rotations. The absence of the competition between the FE mode and $a^-a^-a^-$ tilt in the small rotation amplitude region can be explained as follows: first, the FE mode is mainly dominated by the B-site Ti$^{4+}$ ion when the tilt is weak as can be seen from the computed phonon eigenvectors; second the B-site dominant FE instability is enhanced by tilt since the tilt increases the Ti-O distances (note that the cell is kept fixed in a phonon calculation). When the tilt is large, the A-site contribution to the FE instability becomes larger, and the steric effect enhances the A-site dominant FE instability, as we discuss in the manuscript in detail.

We also consider the $a^0a^0c^-$ tilt case since the ground state structure of bulk SrTiO$_3$ adopts this tilt pattern. For comparison, we also investigate the case of the $a^0a^0c^-$ tilt in CaTiO$_3$. Figures S3 and S4 show that the frequency of the out-of-plane FE mode always increases with the rotation amplitude of the $a^0a^0c^-$ tilt in both CaTiO$_3$ and SrTiO$_3$. On the other hand, the frequency of the in-plane FE mode first increases with the rotation amplitude in CaTiO$_3$, then decreases with the rotation amplitude, similar to the ZnSnO$_3$ case. In contrast, the frequency of the in-plane FE mode always decreases with the rotation amplitude in SrTiO$_3$ because the B-site Ti$^{4+}$ ion is mainly responsible for the FE instability there. To summarize this part, if the B-site has no FE instability (e.g., CaTiO$_3$), there exists a dual coupling between the FE mode and rotation, similar to that in $ZnSnO_3$. Otherwise, the B-site FE instability enhances the cooperative coupling between the FE mode and rotation in some cases [i.e., (i) FE mode along [111] and $a^-a^-a^-$ tilt; (ii) in-plane FE mode and $a^0a^0c^-$ tilt].

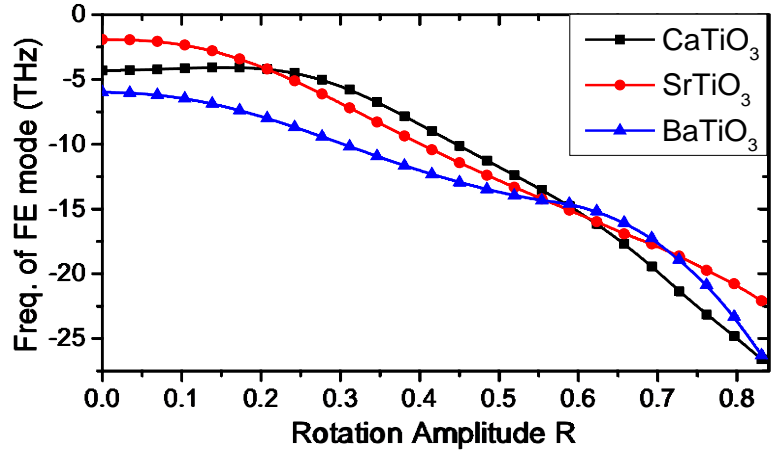

Figure S2. Dependence of the frequency of the (unstable) FE mode on the rotation mode in the R-3c phases of $ATiO_3$ (A= Ca, Sr, Ba).

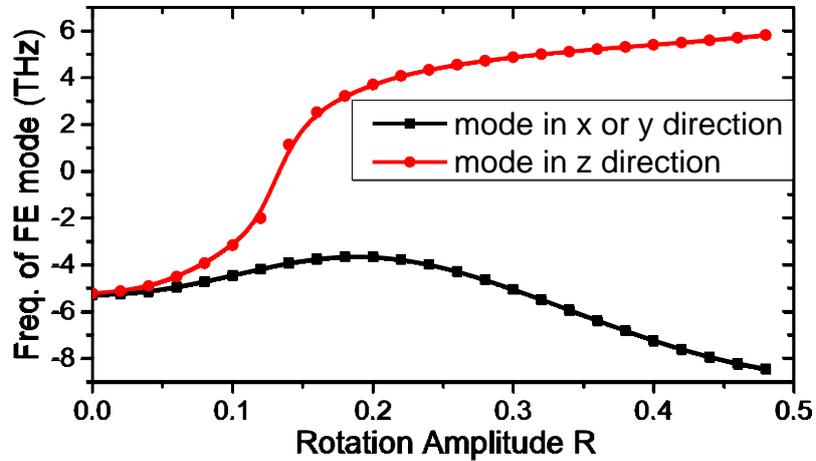

Figure S3. The dependence of the frequency of the soft FE modes on the rotation mode ($a^0a^0a^-$) in the I4/mcm phase of $CaTiO_3$.

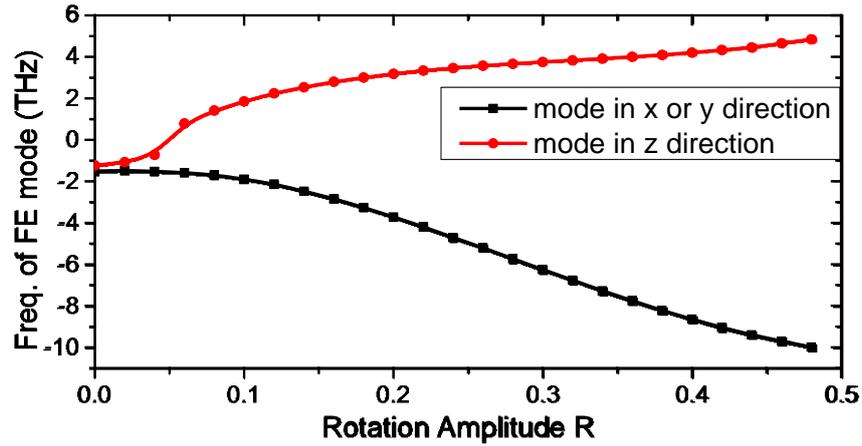

Figure S4. The dependence of the frequency of the soft FE modes on the rotation mode ($a^0a^0a^-$) in the I4/mcm phase of $SrTiO_3$.

## 4. Goldschmidt tolerance factors for the perovskite oxides studied in this work

Table S1: Goldschmidt tolerance factors for some perovskite oxides. We use the ``Crystal Radius'' listed in http://abulafia.mt.ic.ac.uk/shannon/ptable.php. The coordination numbers of A, B and O atoms are chosen tobe eight, six and two, respectively.

| $ABO_3$ | $BaTiO_3$ | $SrTiO_3$ | $CaTiO_3$ | $BiFeO_3$ | $LiOsO_3$ | $InFeO_3$ | $ScFeO_3$ | $ZnSnO_3$ |
|---|---|---|---|---|---|---|---|---|
| **t value** | 1.00 | 0.94 | 0.89 | 0.89 | 0.83 | 0.80 | 0.79 | 0.78 |

## 5. Dual coupling between FE and rotation modes in other $ABO_3$ compounds

In order to further validate the dual nature of thecoupling between AFD and FE modes in other R3c perovskites, DFT calculations similar to those conduced on $ZnSnO_3$ are performed on other R3c systems.We find that the dual nature of thecoupling is not only seen in other small-tolerance-factor insulating perovskites ($ScFeO_3$ and $InFeO_3$), but also occurs in the $BiFeO_3$ system that has a relative large tolerance factor. Furthermore, this dual nature is also present in the "metallic ferroelectric" $LiOsO_3$ that exhibits a small tolerance factor. This suggests that the dual nature of thecoupling between the rotational and FE modes (which arises from an ionic size effect) is a general phenomenon.

The differences in the dependence of the FE mode on the rotation amplitude in these investigated other systems, for the three different relaxation schemes of Fig. 2a of the main manuscript, are worthwhile to mention here. For $BiFeO_3$, when the energy reaches a minimum, the rotation amplitude is small (~ 0.25). Thus, the rotation amplitude at the minimum of the energy locates at the region where the rotation mode tends to suppress the FE mode (see Fig. S5). In contrast, in $ScFeO_3$ and $InFeO_3$, the rotation amplitude at the minimum of the energy is larger than the rotation amplitude corresponding to the minimum of the FE mode. As a result, the rotation amplitude at the minimum of the energy locates at the region where the rotation mode enhances the FE mode (see Figs. S6 and S7). Finally, the case of $LiOsO_3$ (shown in Fig. S8) is a bit different from all the above three cases, but similar to the case of $ZnSnO_3$: when the cell volume is not allowed to change (``fix cell'' case), the rotation amplitude at the minimum of the energy locates at the region where the rotation mode suppresses the FE mode; otherwise, for the other two relaxation schemes, the rotation amplitude at the minimum of the energy locates at the region where the rotational mode enhances the magnitude of the FE mode.

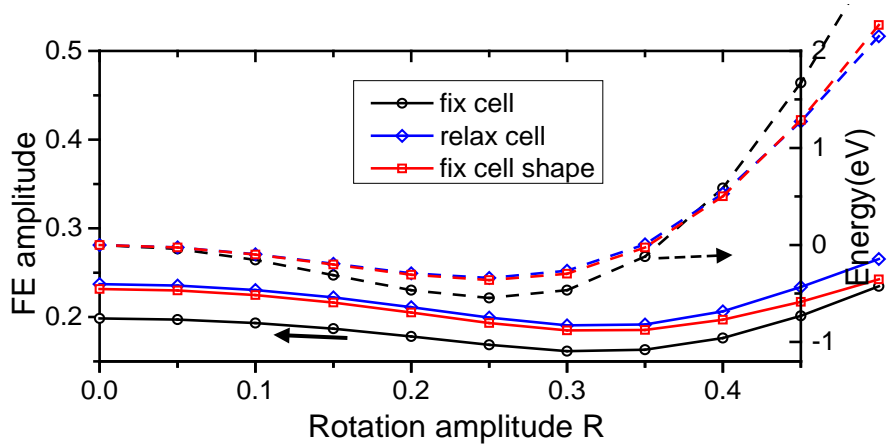

Figure S5. Total energy (right) and FE amplitude (left) versus the rotation mode in R3c $BiFeO_3$. The G-type antiferromagnetic state is adopted in the calculations, and the three different structural relaxation strategies indicated in the manuscript (i.e., ``fix cell'', ``fix cell shape'' and ``relax cell'') are considered. Total energy is in eV/10-atoms.

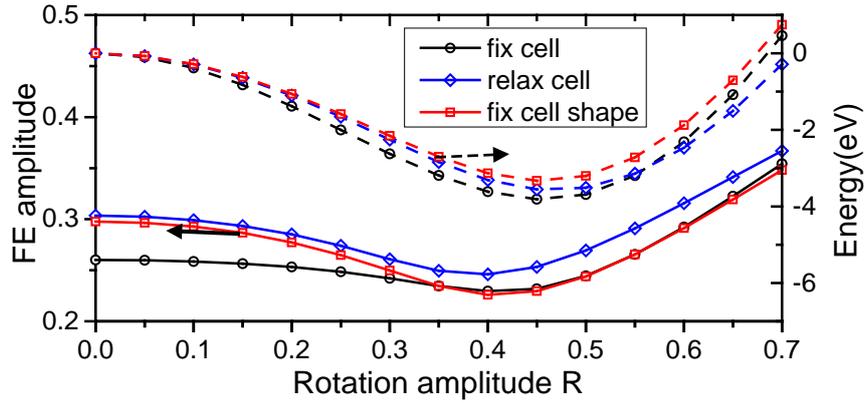

Figure S6. Same as Fig. S5 but for the G-type antiferromagnetic R3c ScFeO$_3$.

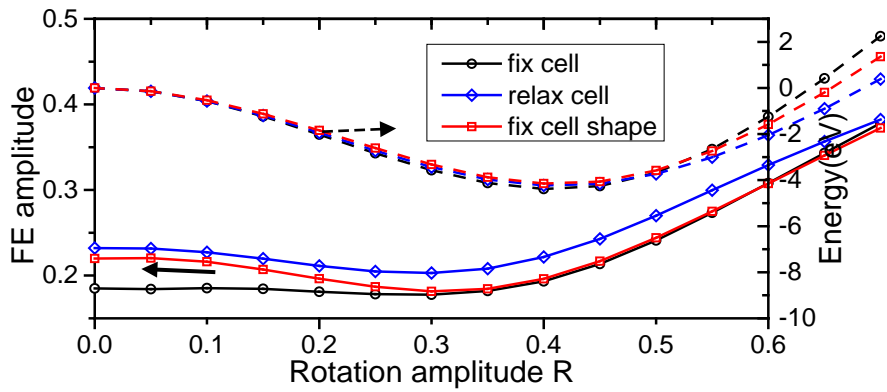

Figure S7. Same as Fig. S5 but for the G-type antiferromagnetic R3c InFeO$_3$.

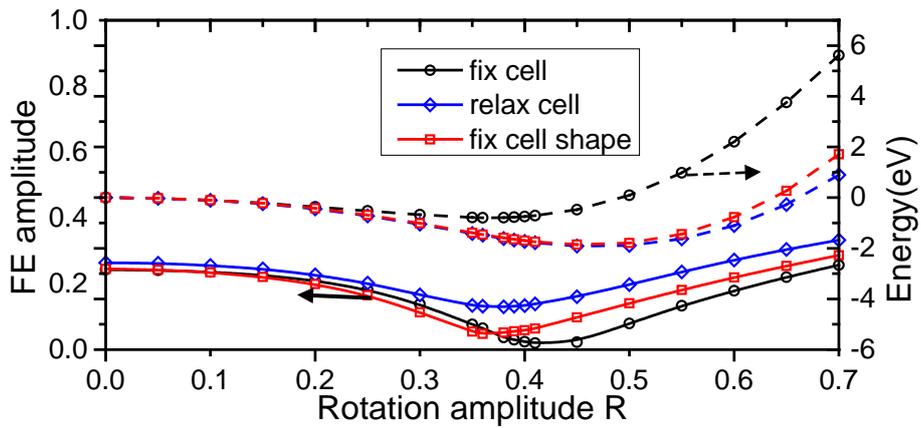

Figure S8. Same as Fig. S5 but for R3c LiOsO$_3$.

## 6. Self-force constants computed with a Lennard-Jones like model potential

In order to validate the size effect in the manuscript, we examine the interaction form between Zn and O ions by calculating the energy of the Zn-O dimer as a function of Zn-O distance with the DFT [15,16] PBE [5]functional. In our calculations, we put the Zn-O dimer in a large supercell (the cell parameters are a = b = c = 20 Å), and then change the Zn-O distance. We find that the interaction between the Zn and O ions cannot be well described by the simple Lennard-Jones (LJ) model potential $V(r) = A/r^{12} - B/r^6$ but rather one needs to consider a modified LJ-like model potential of the form $V(r) = A/r^m - B/r^n$ with the fitted parameters A, B, m and n being 175, 136, 4.5 and 3.6 (energy in eV, distance in Å), respectively, to mimic well such interaction (see Fig. S9). With this modified LJ-like model potential, we find that the $\Phi^{self}_{\|[111],\|[111]}$ self-force constant of the Zn ion first increases but then decreases with the octahedral rotation (inset of Fig. 3a of the main manuscript), in agreement with the DFT results displayed in Fig. 3a of the main manuscript.

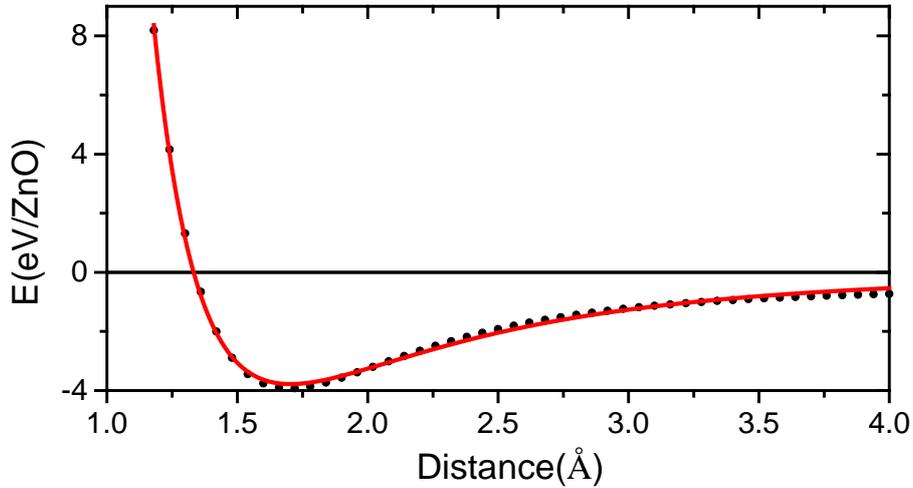

Figure S9. The energy of the Zn-O dimer as a function of the Zn-O distance. The black dots are the calculated DFT data and the red curve is the fit by the modified LJ-like potential.

7. **Dependence of the frequency of the FE modes on the rotation amplitude of $a^0a^0c^-$ or $a^0a^0c^+$ tilts in ZnSnO$_3$**

We also examined the dependence of the frequency of the FE modes on the rotation modes in ZnSnO$_3$ with tilts different from the a$^-$a$^-$a$^-$ tilt pattern. For the antiphase rotation along the [001] axis (a$^0$a$^0$c$^-$tilt) (see Fig. S10) or in-phase rotation along the [001] axis (a$^0$a$^0$c$^+$tilt) (see Fig. S11), the frequency of the FE mode along the [001] axis always increases with the rotation amplitude. The case of in-plane FE modes is different: when the rotation amplitude is small, the frequency of the in-plane FE modes also increases with the rotation amplitude, while it starts to decrease for very large rotation amplitude (i.e., when R > 0.45). This suggests that there is also a dual nature for the coupling between the in-plane FE modes and the a$^0$a$^0$c$^-$ or a$^0$a$^0$c$^+$ tilt mode, as similar to the a$^-$a$^-$a$^-$ tilt case.

Similar to the a$^-$a$^-$a$^-$ tilt case, the twelve NN oxygen ions are also separated into three groups (see Fig. S12) for the a$^0$a$^0$c$^-$ and a$^0$a$^0$c$^+$ cases: (1) four in-plane oxygen ions that do not change positions with tiltings so that the distances between these in-plane oxygen ions and the Zn ion remain unchanged with rotation; (2) among the eight out-of-plane oxygen ions, four of them move away from the Zn ion while the other four out-of-plane oxygen ions move closer to the Zn ion so that the Zn-O distances decrease with the rotation amplitude R. If the Zn ion undergoes in-plane displacements, the shortest distances between the Zn ion and the four out-of-plane oxygen ions will be increased so that the Pauli repulsion will be decreased. Such a result is very similar to that of the a$^-$a$^-$a$^-$ tilt case, suggesting that the dual nature of the coupling between FE and rotation modes has a simple steric origin and is applicable to any tilt case.

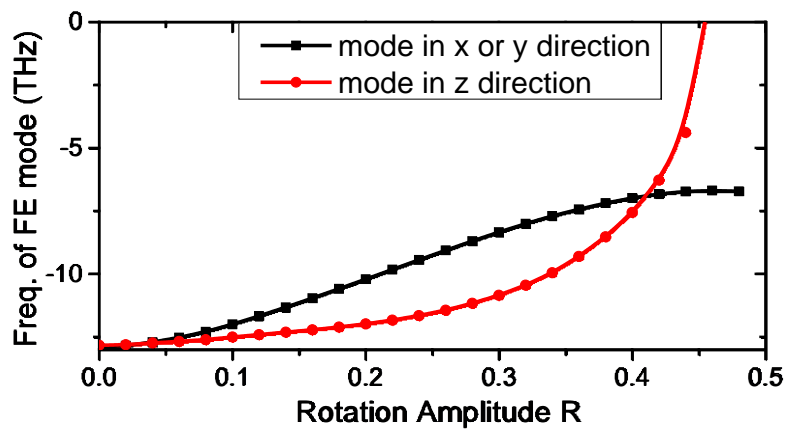

Fig. S10. Dependence of the frequency of the (unstable) FE modes on the rotation

mode ($a^0a^0c^-$) in the I4/mcm phase of ZnSnO$_3$.

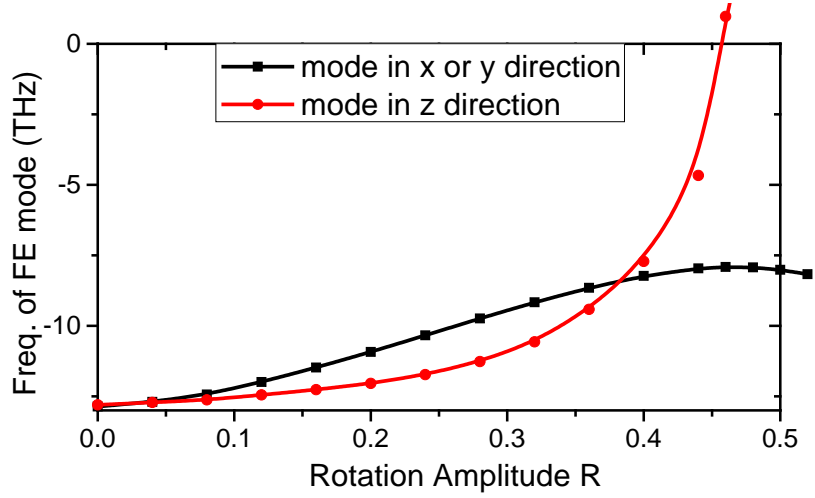

Fig. S11. Dependence of the frequency of the (unstable) FE modes on the rotation mode ($a^0a^0c^+$) in the P4/mbm phase of ZnSnO$_3$.

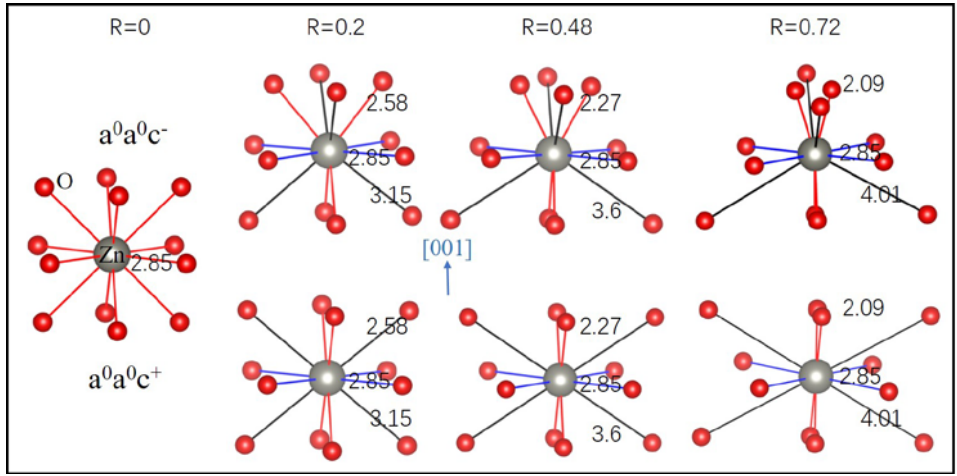

Fig. S12. Local environment around the Zn ion for four different rotation amplitudes (R = 0, 0.2, 0.48 and 0.72) for the $a^0a^0c^-$ and $a^0a^0c^+$ cases. The Zn-O distances (in Å) are indicated.

## 8. Comparative study on the FE-AFD coupling in SrTiO$_3$

In Aschauer and Spaldin' work[9], a strain-mediated FE—AFD cooperation was reported. Here we will examine to see to what extent they are compatible with our present investigation. Our results lead to the following conclusions: (1) In agreement with what was observed in Ref. [9], we find that (see Figure S4), in absence of a tetragonal cell distortion, the frequency of the z-polarized FE mode increases as a function of the $a^0a^0c^-$ tilting, never reaching the "cooperative regime". Further calculations (not shown here) confirm that, as found in Ref. [9], the cooperation between parallel FE and AFD modes requires the tilt-induced *strain relaxation* of the cell. (2) In the case of the xy-polarized FE modes, we find an obvious cooperation with the $a^0a^0c^-$ tilt (see Figure S4). This case – which was not considered in the work of Aschauer and Spaldin as far as we can tell – shows that the FE—AFD cooperation is possible in SrTiO$_3$ even in absence of strain relaxation. This is because the $a^0a^0c^-$ tilt results in longer in-plane Ti-O distances, thus favoring the in-plane FE instability. (3) Finally, for the $a^-a^-a^-$ tilts and parallel FE modes, the coupling is always cooperative as we discussed above (see Figure S2). For comparison, we also examine the coupling between the FE modes and tilts other than the $a^-a^-a^-$ pattern in ZnSnO$_3$. In the $a^0a^0c^-$ and $a^0a^0c^+$ tilt cases (see Figs. S10 and S11), the out-of-plane FE mode is always suppressed by the tilt, while there is a dual coupling between the in-plane FE modes and $a^0a^0c^-$ tilt mode. All these results can be understood with the steric argument given in section 5, which is similar to the $a^-a^-a^-$ tilt case discussed in the main manuscript.

## 9. Pressure behavior of the electrical polarization in R3c ZnSnO$_3$

It is well-known that hydrostatic pressure typically weakens or even suppresses electrical polarization in most perovskites[17,18], since the short-range repulsions (favoring a paraelectric state) increase more rapidly thanlong-range interactions (preferring a FE state) as pressure increases. On the other hand, improper ferroelectricity in hexagonal ReMnO$_3$ and ReFeO$_3$ (where Re stands for a rare-earth element) was found to be stronger under compression[19,20], as a result of the concomitant enhancement of the primary order parameter, that is, the so-called *K$_3$* AFD tilting mode. In view of such an interesting behavior, one may wonder how pressure will affect ferroelectricity in LN-type compounds. Indeed, while being proper in nature, we find that ferroelectricity in materials like ZnSnO$_3$ is linked to the amplitude of the non-polar AFD distortion, and the FE response to any external perturbation should thus be conditioned by the response of the tilting modes. To address such an issue, the FE polarization and AFD rotation amplitudes are showed in Fig. S13 as a function of hydrostatic pressure up to 80 kbar, in the R3c state of $ZnSnO_3$. Three different ways to relax the structures are adopted here: (1) both atomic positions and lattice vectors are fully relaxed, which is denoted as the ``full relax'' case in Fig. S13a; (2) the lattice vectors are optimized and atomic positions are relaxed, except those related to the AFD rotation, which are fixed to be equal to their zero-pressure values; this situation is termed ``fix rotation'' here; and (3) the cell shape is taken to be cubic, while the atomic positions and the volume of such a cubic cell are allowed to relax in order to minimize the energy; such a case is denoted ``fix cell shape''. In all these three cases, Fig. S13 indicates that the electrical polarization increases with the hydrostatic pressure, which constitutes a rather novel effect for proper ferroelectrics. When no rhombohedral distortion is allowed (i.e., in the ``fix cell shape'' scenario), the magnitude of the FE polarization is smaller than in the other two cases, but the polarization-*versus*-pressure curve has a similar slope. Note that we also found that the octahedral rotation amplitude increases with pressure in the ``full relax case''; however, such an increase is rather small, which explains why the electric polarization is almost identical in the ``full relax'' and ``fix rotation'' cases.

Let us now try to understand why polarization is enhanced by hydrostatic pressure in the R3c phase of $ZnSnO_3$. For that, we resort to our Landau model detailed in the manuscript, in general, and to some of its extracted parameters (from the ``fix cell shape'' case), in particular. For instance, Fig. S13(b) shows that $A_{u^2}$ and $A_{u^2R^2}$ both increase when decreasing the lattice constant of the cubic cell, i.e. when increasing the pressure; such results reflect mechanisms I and II of Fig. S13(c), respectively. This is in agreement with the fact that pressure typically suppresses polarization in proper ferroelectrics. On the other hand, $A_{u^2R^4}$ quickly decreases when reducing the lattice constant (mechanism III in Fig. S13(c)), which therefore constitutes the main and novel mechanism responsible for the unusual enhancement of polarization under compression. [One could also imagine that pressure might change the rotation amplitude R (mechanism IV in Fig. S13(c)) and thus indirectly the polarization, to explain the pressure-dependent results. However, such a scenario is not occurring here because, e.g., our ``fixed rotation'' simulations shown in Fig. S13(a) show a similar increase of polarization with pressure than in the ``full relax'' case.]

The dependence of $A_{u^2R^4}$ with lattice constant can be understood in the following intuitive way: for a given AFD amplitude, a smaller lattice constant naturally corresponds to a smaller distance between the Zn ion and its three NN oxygens, resulting in a stronger Pauli repulsion between Zn and O ions and thus an enhanced FE instability. Note that this is consistent with Fig. 3 of the main manuscript and analysis based on a simple Lennard-Jones-like model.

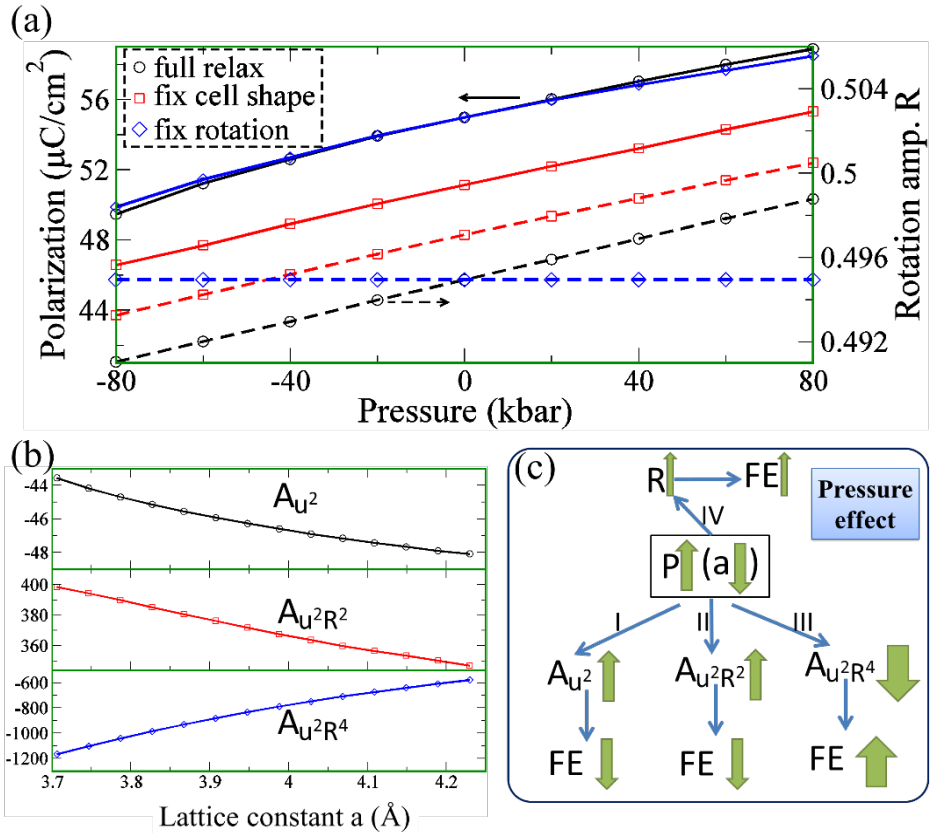

Figure S13. Pressure behavior of the electrical polarization in R3c $ZnSnO_3$. (a) Amplitudes of the polarization (solid lines) and rotation (dashed lines) as a function of pressure from DFT calculations. Three different ways for structural optimization are adopted, namely "fix cell shape", "full relax", and "fix rotation" (see text). (b) Coefficients ($A_{u^2}$, $A_{u^2R^2}$ and $A_{u^2R^4}$) related to the second order $u^2$ terms as a function of the cubic lattice constant. (c) Schematization of four different mechanisms for the pressure effect on the FE polarization. In ZnSnO$_3$, $A_{u^2R^4}$ decreases quickly with pressure, which is mainly responsible for the enhancement of FE polarization by pressure.

**References**

[1] P. E. Blöchl, Physical Review B **50**, 17953 (1994).
[2] G. Kresse and D. Joubert, Physical Review B **59**, 1758 (1999).
[3] G. Kresse and J. Furthmüller, Physical Review B **54**, 11169 (1996).
[4] G. Kresse and J. Furthmüller, Computational Materials Science **6**, 15 (1996).
[5] J. P. Perdew, K. Burke, and M. Ernzerhof, Physical Review Letters **77**, 3865 (1996).
[6] R. D. King-Smith and D. Vanderbilt, Physical Review B **47**, 1651 (1993).
[7] D. Vanderbilt and R. D. King-Smith, Physical Review B **48**, 4442 (1993).
[8] R. Resta, Reviews of Modern Physics **66**, 899 (1994).
[9] U. Aschauer and N. A. Spaldin, Journal of Physics: Condensed Matter **26**, 122203 (2014).
[10] J. P. Perdew, A. Ruzsinszky, G. I. Csonka, O. A. Vydrov, G. E. Scuseria, L. A. Constantin, X. Zhou, and K. Burke, Physical Review Letters **100**, 136406 (2008).
[11] K. Parlinski, Z. Q. Li, and Y. Kawazoe, Physical Review Letters **78**, 4063 (1997).
[12] A. Togo, F. Oba, and I. Tanaka, Physical Review B **78**, 134106 (2008).
[13] R. A. Cowley, Physical Review **134**, A981 (1964).
[14] L. Bellaiche and J. Íñiguez, Physical Review B **88**, 014104 (2013).
[15] P. Hohenberg and W. Kohn, Physical Review **136**, B864 (1964).
[16] W. Kohn and L. J. Sham, Physical Review **140**, A1133 (1965).
[17] E. Bousquet and P. Ghosez, Physical Review B **74**, 180101 (2006).
[18] G. A. Samara, T. Sakudo, and K. Yoshimitsu, Physical Review Letters **35**, 1767 (1975).
[19] C. Xu, Y. Yang, S. Wang, W. Duan, B. Gu, and L. Bellaiche, Physical Review B **89**, 205122 (2014).
[20] H. Tan, C. Xu, M. Li, S. Wang, B.-L. Gu, and W. Duan, Journal of Physics: Condensed Matter **28**, 126002 (2016).